\documentclass[english,aps, prl]{revtex4}
\usepackage[T1]{fontenc}
\usepackage[latin9]{inputenc}
\usepackage{amssymb}
\usepackage{graphicx}

\makeatletter

\usepackage{newfloat}
\DeclareFloatingEnvironment[name={Supplementary Figure}]{suppfigure}

\providecommand{\tabularnewline}{\\}

\@ifundefined{textcolor}{}
{%
 \definecolor{BLACK}{gray}{0}
 \definecolor{WHITE}{gray}{1}
 \definecolor{RED}{rgb}{1,0,0}
 \definecolor{GREEN}{rgb}{0,1,0}
 \definecolor{BLUE}{rgb}{0,0,1}
 \definecolor{CYAN}{cmyk}{1,0,0,0}
 \definecolor{MAGENTA}{cmyk}{0,1,0,0}
 \definecolor{YELLOW}{cmyk}{0,0,1,0}
}

\@ifundefined{showcaptionsetup}{}{%
 \PassOptionsToPackage{caption=false}{subfig}}
\usepackage{subfig}
\makeatother

\usepackage{babel}
\begin{document}

\title{Nanoscale structuring of tungsten tip yields most coherent electron point-source}

\author{Josh Y. Mutus}

\affiliation{Department of Physics, University of Alberta, 11322-89 Avenue, Edmonton,
Alberta, T6G 2G7, Canada}

\affiliation{National Institute for Nanotechnology, National Research Council
of Canada, 11421 Saskatchewan Drive, Edmonton, Alberta, T6G 2M9, Canada }

\author{Lucian Livadaru}
\affiliation{Department of Physics, University of Alberta, 11322-89 Avenue, Edmonton,
Alberta, T6G 2G7, Canada}

\affiliation{National Institute for Nanotechnology, National Research Council
of Canada, 11421 Saskatchewan Drive, Edmonton, Alberta, T6G 2M9, Canada }

\author{Radovan Urban}
\affiliation{Department of Physics, University of Alberta, 11322-89 Avenue, Edmonton,
Alberta, T6G 2G7, Canada}

\affiliation{National Institute for Nanotechnology, National Research Council
of Canada, 11421 Saskatchewan Drive, Edmonton, Alberta, T6G 2M9, Canada }

\author{Jason Pitters}

\affiliation{Department of Physics, University of Alberta, 11322-89 Avenue, Edmonton,
Alberta, T6G 2G7, Canada}

\affiliation{National Institute for Nanotechnology, National Research Council
of Canada, 11421 Saskatchewan Drive, Edmonton, Alberta, T6G 2M9, Canada }

\author{A. Peter Legg}
\affiliation{Department of Physics, University of Alberta, 11322-89 Avenue, Edmonton,
Alberta, T6G 2G7, Canada}

\affiliation{National Institute for Nanotechnology, National Research Council
of Canada, 11421 Saskatchewan Drive, Edmonton, Alberta, T6G 2M9, Canada }

\author{Mark H. Salomons}
\affiliation{Department of Physics, University of Alberta, 11322-89 Avenue, Edmonton,
Alberta, T6G 2G7, Canada}

\affiliation{National Institute for Nanotechnology, National Research Council
of Canada, 11421 Saskatchewan Drive, Edmonton, Alberta, T6G 2M9, Canada }

\author{Martin Cloutier}

\affiliation{Department of Physics, University of Alberta, 11322-89 Avenue, Edmonton,
Alberta, T6G 2G7, Canada}

\affiliation{National Institute for Nanotechnology, National Research Council
of Canada, 11421 Saskatchewan Drive, Edmonton, Alberta, T6G 2M9, Canada }

\author{Robert A. Wolkow}

\email{rwolkow@ualberta.ca}

\affiliation{Department of Physics, University of Alberta, 11322-89 Avenue, Edmonton,
Alberta, T6G 2G7, Canada}

\affiliation{National Institute for Nanotechnology, National Research Council
of Canada, 11421 Saskatchewan Drive, Edmonton, Alberta, T6G 2M9, Canada }

\begin{abstract}
This report demonstrates the most spatially-coherent electron source 
ever reported. A coherence angle of 14.3 $\pm 0.5^{\circ}$ 
was measured, indicating a virtual source size of 1.7 $\pm 0.6$ \AA\ using an extraction voltage of 89.5 V. The nanotips under study were crafted using a spatially-confined, field-assisted nitrogen etch which removes material from the periphery of the tip
apex resulting in a sharp, tungsten-nitride stabilized, high-aspect
ratio source. The coherence properties are deduced from holographic
measurements in a low-energy electron point source microscope with
a carbon nanotube bundle as sample. Using the virtual source size
and emission current the brightness normalized to 100 kV is found
to be 7.9$\times10{}^{8}$ A/sr cm$^{2}$.
\end{abstract}
\maketitle


Electron sources have had a substantial impact on modern physics,
since the invention of the cathode-ray tube well over a century ago.
Now, electron sources are ubiquitous in science and industry in a variety
of areas such as spectroscopy, microscopy and lithography. There have
been sequential innovations in the area of nanotip fabrication and
characterization \cite{fink1986mono,fink1988point,chang2009fully,cho2004quantitative}.

Nanotips are of interest as field-emission electron sources as well as gas field ion sources
due to their high brightness and coherence \cite{pitters2012creation,urban2012gas,unnanotip3}.
A high degree of coherence of such sources enables many
basic applications in electron interferometry and holography, relying
on splitting and self-interference of single-particle amplitudes.
Essentially, these types of experiments exploit and probe first-order coherence
properties, typically manifest as fringes in an interference pattern,
or topological effects such as in the Ahronov-Bohm effect.

While these aspects of quantum mechanical behavior of electrons
are well established, more subtle manifestations of quantum mechanics,
such as those derived from second-order coherence properties
of electrons and, more generally, of fermions
still await detailed exploration. In this sense, the most formidable
experimental challenge resides in producing sufficiently bright and coherent
(therefore degenerate in the quantum mechanical sense) 
ensembles of fermions with well-understood interactions.
A beam of freely propagating electrons qualifies as one of the best candidates
for this description \cite{silverman2008quantum}. 
The beam character of this type of ensemble is particularly 
amenable to a great variety of experimental setups with correlated 
and entangled fermions from the Hanbury-Brown
and Twiss experiment to the Ahronov-Bohm effect, and combinations thereof
\cite{silverman2008quantum}.
Therefore, developing more degenerate beams,  such as those obtained 
from crafting ultrasharp metallic tips at the nanoscale, will enable among 
other things, a better understanding of correlated particle behavior
in vacuum or in the presence of fields, a mostly uncharted experimental area.
The enhanced brightness of such electron beams also open the possibility
of probing ultrafast dynamics \cite{barwick2007laser}.

Arguably, with the availability of a highly coherent source, the most immediate 
application would  be holographic imaging with electrons, as first recognized by Gabor \cite{gabor1948new}.
Specifically, the degree of spatial coherence can be shown to be the critical 
limiting factor in the resolution of point-source electron holography \cite{stevens2011individual,stevens2009resolving}\footnote{Degeneracy, coherence and brightness of a point source are all inextricably related as they all depends on the virtual source size. Thus, minimizing the latter is key to maximizing the related quantities.}.
As it is difficult, if not impossible, to envision an electron source with a greater spatial
coherence than that generated by field emission from a single atom,
such emitters are now being explored extensively.

The development of techniques for creating nanotips that emit electrons
from a single terminal atom have been ongoing since the first work
by Fink in 1986 \cite{fink1988point,fink1986mono}. Recently, the
latter have been shown to be completely coherent: the emission across
the entire opening-angle shares one phase \cite{chang2009fully}.
However, although completely coherent, the emission angle for such
tips remains rather small, typically a few degrees \cite{chang2009fully}.
Consequently, the resulting spatial coherence of such sources has been limited by
low beam divergence angles. 
Here we show that our single-atom tip not only has the property of 
being completely coherent, but also exhibits a coherence angle much larger 
than measured previously, and indeed than previously thought feasible \cite{stevens2009resolving}.
This implies, among other things, a much greater resolution 
in the point-source holographic imaging with such a tip.

The degree of spatial coherence of an electron point source can be experimentally determined
using a low-energy electron point source (LEEPS) microscope \cite{spence1994electron}, 
a method that has been well established. The state of the art in this technique
was recently described by Hwang et al.\cite{tsong2013}

The LEEPS microscope consists of
only a source, a sample and a detector. A bias is placed between a sharp metal 
tip and a grounded sample, causing electrons to be field emitted from the tip, 
past the sample and toward the detector. The coherent electron wave emitted 
from the tip is partially  scattered off the sample, while the rest propagates 
towards the detector. These waves interfere at the detector to create a hologram. 
The width of the hologram is a measure of the angular-width of the coherent wavefront emitted by
the tip. As in any microscope, the numerical aperture dictates the
resolution. In a LEEPS microscope, the coherence angle, $\alpha$,
is dictated by the character of the source. However, it is limited
by many attributes of the microscope itself: uncontrolled fields and
mechanical and electrical stability of the instrument can also prevent
attainment of high brightness and coherence. The diffraction limited resolution,
$R$, of the LEEPS microscope is given by:

\begin{equation}
R\geq\lambda/2\sin\alpha,\label{eq:diff_limit}
\end{equation}
where $\lambda$ is the wavelength of the incident electrons. Another
widely used estimate of resolution is the virtual source size, given
by \cite{spence1994electron}:

\begin{equation}
R_{v}\thickapprox\lambda/\pi\alpha.\label{eq:vss}
\end{equation}

In this paper we report on the coherence properties 
\footnote{While many published results in this area have put measurements in terms
of half-width-at-half-maximum, in this work we quote the coherence
angle as the half-width of the entire pattern. This enables us to
put our measurements easily in context of the Numerical Aperture (NA)
of the microscope since each fringe present in the image present in
the pattern will contribute to the overall resolution of the reconstructed
hologram. A comparison of our work to landmark works in the literature
can be found in Table \ref{tab:comp} contained in the Supplementary
Information.} of a nanotip made by removing material from the shank of the tip, rather than building
upon it \cite{pitters2013tip}. This technique uses a field-assisted, spatially-confined
nitrogen etch and was reported in 2006 \cite{pitters2006tungsten}.
Such tips are not only terminated by a single tungsten atom but are
uniquely stable due to the nitrogen-rich layer formed during fabrication,
which coats the surface area surrounding the apex. This greatly inhibits
metal diffusion and allows the tip to survive exposure to atmosphere
and heating to a thousand degrees Celsius. A detailed description
of the tip etching procedure can be found in the Supplementary Information 

\begin{figure}
\captionsetup[subfigure]{labelformat=empty,labelsep=colon}   \captionsetup{justification=raggedright, singlelinecheck=false}\subfloat{\includegraphics[width=1\columnwidth]{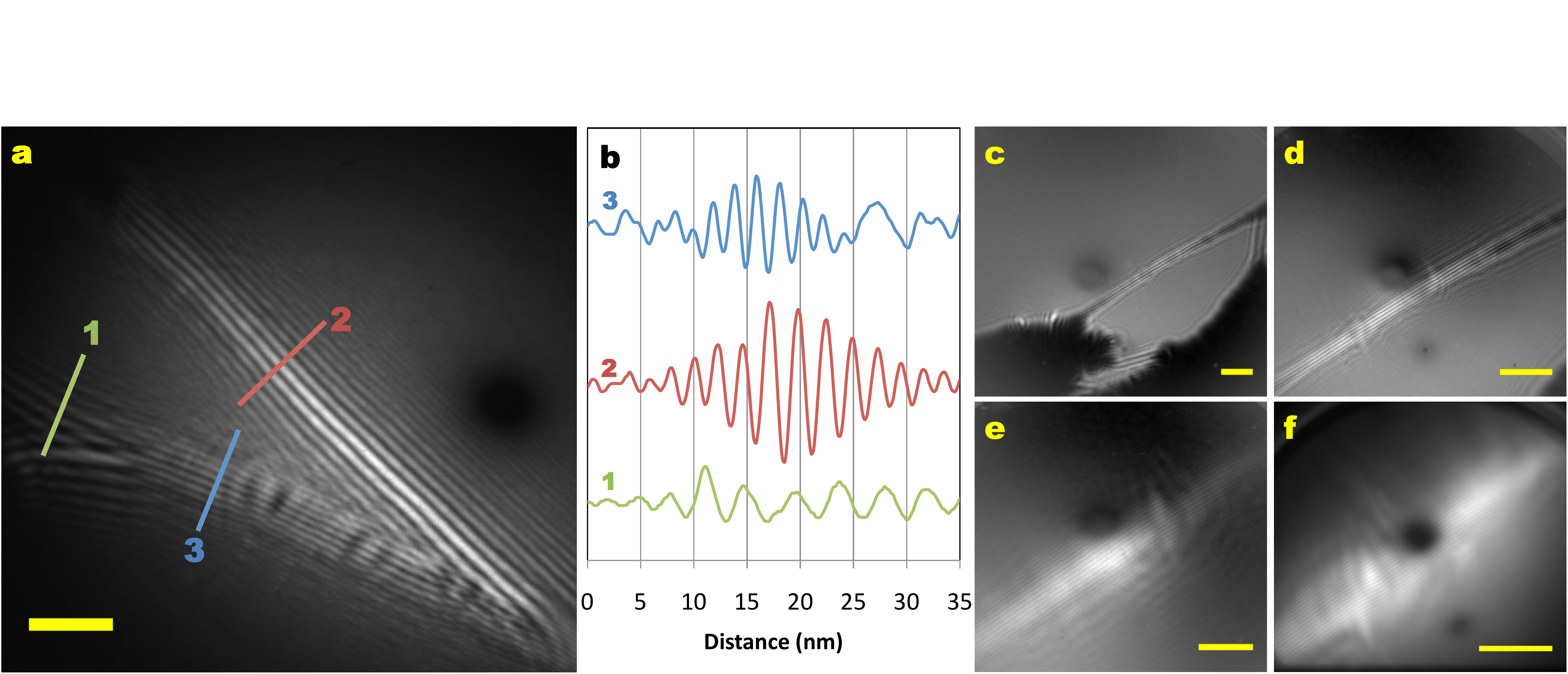}
}

\caption{At low magnification the electron beam forms a broad pattern, analogous to a double slit,
due to simple scattering off the bundle. This is shown in (a) (scale
bar 20nm) and the corresponding line profiles in (b) where a branched
bundle is imaged. (c-f) LEEPS images of carbon nanotube bundles as the tip approaches the
bundle. As the magnification increases,
the pattern fills the emission spot on the screen revealing complex interferometric detail. 
It is important to note that upon magnification fringe patterns are evident
that are progressively finer from (c) to (f).
Scale bars are 100 nm and 50 nm for
c-d and e-f, respectively.\label{fig:leeps_images} Tip voltages and
currents -122 V and 0.28 nA, -87.3 V and 0.29 nA, -89.5 V and 1.2
nA for images (d), (e) and (f) respectively.}
\label{fig:branches}
\end{figure}


\begin{figure}
\includegraphics[width=1\columnwidth]{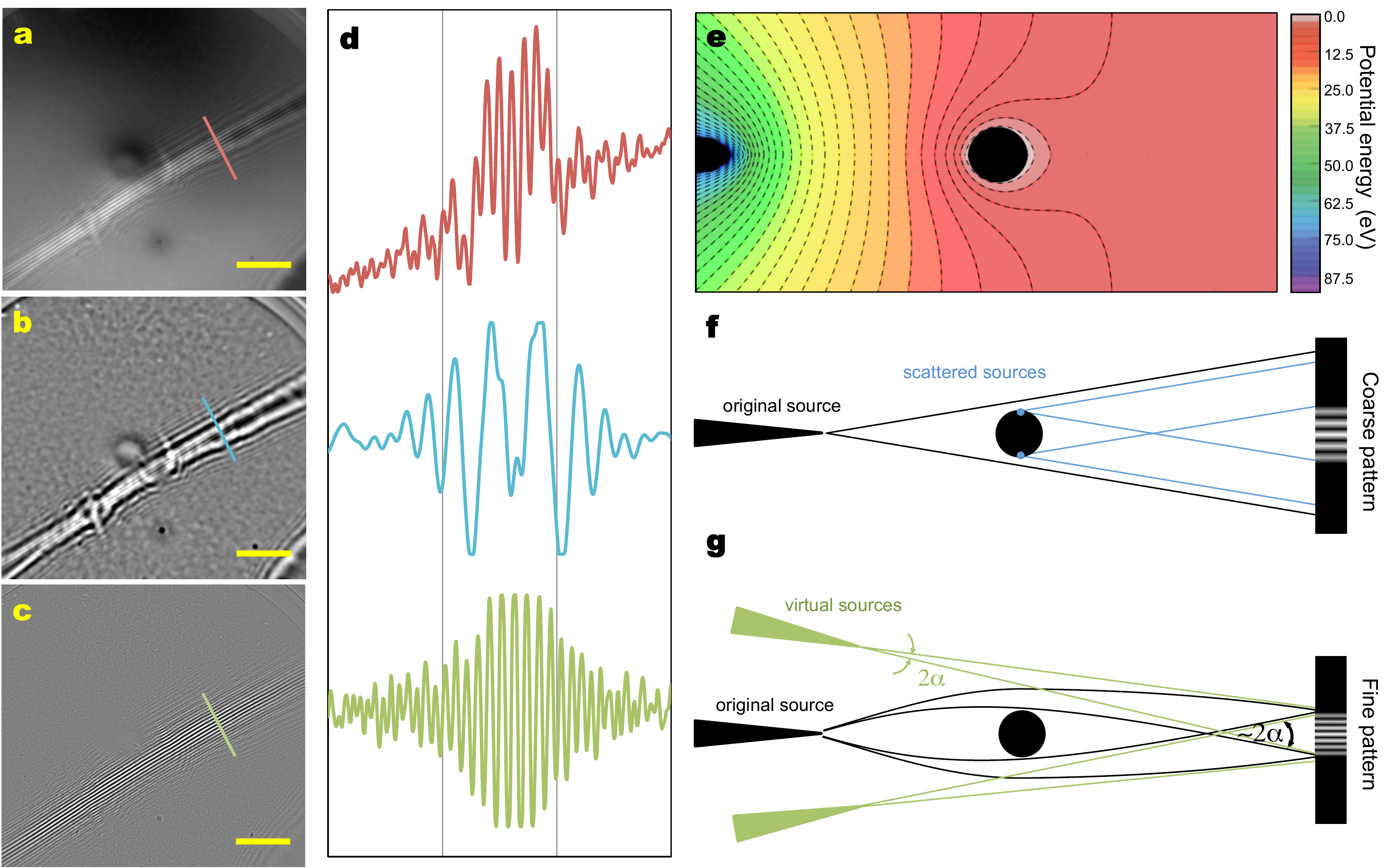}

\captionsetup{justification=raggedright, singlelinecheck=false}

\caption{Filtered versions of Figure \ref{fig:branches}(d) show each fringe
pattern is composed of two sets of fringes of distinct origin. Figures (b) and
(c) are low-pass and high-pass filtered images of (a). Their respective
line profiles are shown in (d). In order to interpret the pattern, both elastic scattering off the bundle and deflection due to the field surrounding the bundle must be considered. Equipotential surfaces between the biased tip and the grounded bundle, as shown in (e) give rise to the electron bi-prism effect. A coarse set of fringes due to scattering of the bundle itself (f) and can be explained as a simple double slit where the edges of the bundle are the sources.  The fine set of fringes apparent in the image is
due to the bi-prism effect shown schematically in (g). The overlap angle calculated here represents a lower bound on the coherence angle of the original beam. The pitch of the finer set of fringes can be tuned by varying the tip-sample distance and voltage as shown in the supplementary information.
  Scale
bar is 100nm.}
\label{fig:regimes}
\end{figure}

\begin{figure*}[t]
\captionsetup[subfigure]{labelformat=empty, labelsep=colon}\captionsetup{justification=raggedright, singlelinecheck=false}\subfloat[\label{fig:coh_anglepic}]{\includegraphics[height=0.26\columnwidth]{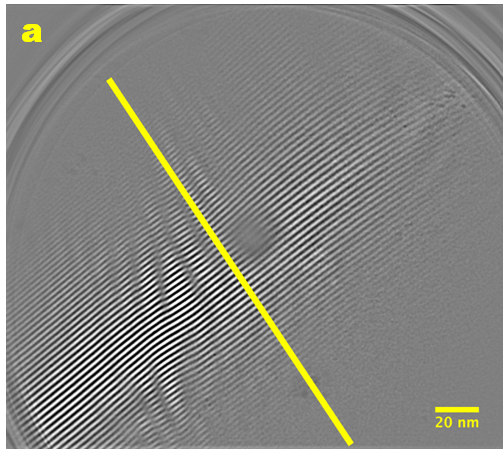}

}\subfloat[\label{fig:coh_angleplot}]{\includegraphics[height=0.26\columnwidth]{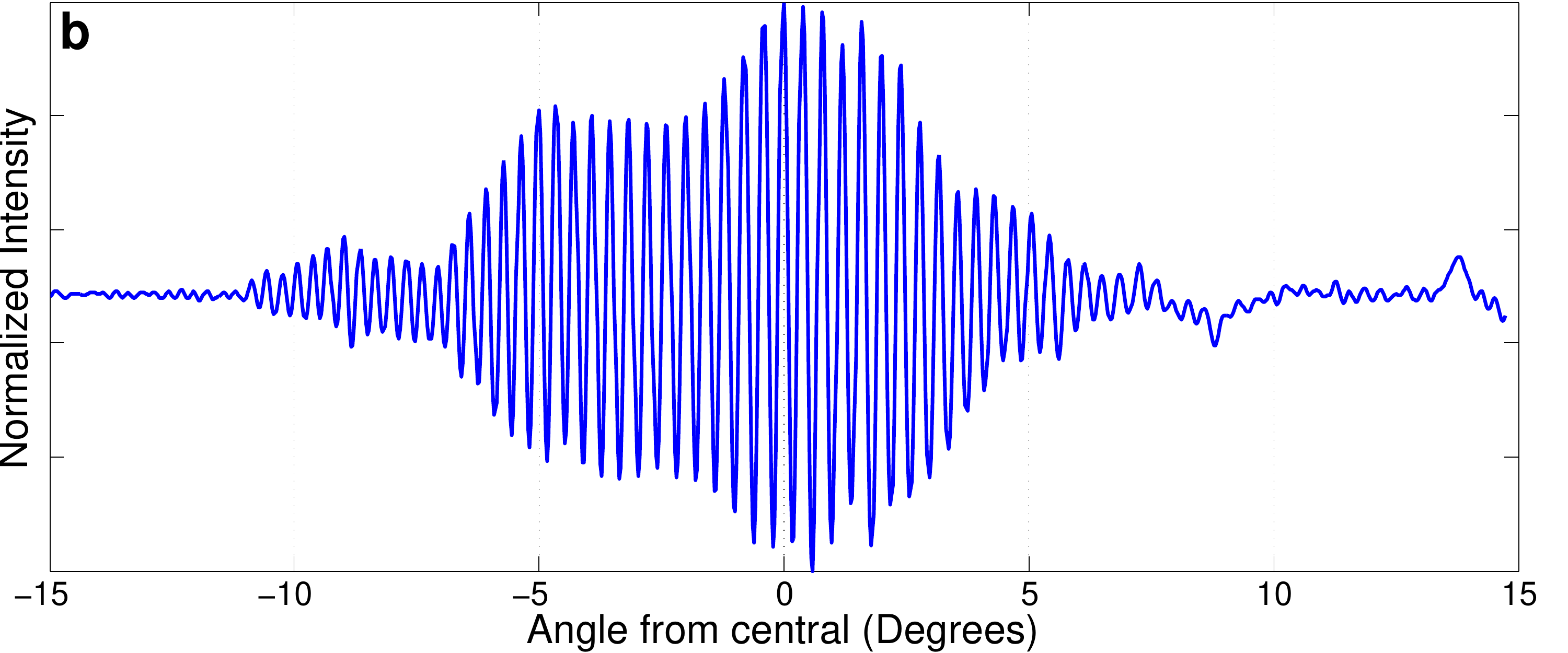}

}

\caption{(a) shows a high-pass filtered version of image \ref{fig:branches}(d)
(scale bar is 50nm). The interference pattern fills the detector with
high-frequency fringes. High order fringes, not obvious in (a) are
visible in the line profile shown in (b). The beam is completely coherent
and possesses a coherence angle greater than 14$^{\circ}$.\label{fig:coh_angle}}
\end{figure*}

The coherence properties of the beam were measured using a custom-designed
LEEPS microscope \cite{mutus2011low}. The details of its construction
are the subject of a future publication. Here we give a brief description:
the tip, sample and motors are contained in the scanning head. The
motors are used to position the tip relative to the sample with subnanometer
precision. The scanning head is suspended above a micro-channel plate
(MCP). A modest, negative bias voltage (from 60-500 V, depending on the
separation between tip and sample) is applied to the tip, with the
sample grounded. Electrons are field-emitted from the tip, past
the sample and toward the MCP some distance away. A magnified image
of the region of the sample irradiated by the electrons is projected
on the MCP. The image is magnified by the ratio of the tip-sample
to the tip-MCP distance. In this work, the tip-MCP distance is 4 cm, but it can be varied easily from 0.5 to 20 cm by moving
both tip and sample relative to the MCP. 

Carbon nanotube (CNT) bundles were suspended across holes in a 50 nm thick
silicon nitride window. These windows were perforated by a regular
array of 2 $\mu$m holes on a 4 $\mu$m pitch \cite{grid}. In order
to make the windows conductive, they were coated in 5 nm Cr and 45
nm Au. These suspended nanotube bundles were brought into the LEEPS
microscope for imaging.

The images generated by the LEEPS microscope showed two distinct
fringe patterns. At low magnifications, diffraction patterns due to
direct scattering off the bundle are observed. Such patterns, analogous to those from a double slit, are shown
in Figure \ref{fig:branches}a and the corresponding profiles in \ref{fig:branches}b. 

At higher magnifications, a finer fringe pattern becomes visible.
Near the central maximum, these fringes are sinusoidally spaced and
reminiscent of those generated by an electron bi-prism. In transmission
electron microscopy (TEM), an electron bi-prism is made from a thin, conductive
filament, biased between two grounded plates in order to create a
beam-deflecting electrostatic field on each side of the 
filament \cite{lichte2008electron}. 
A carbon nanotube bundle approximates this effect in LEEPS by having an electric field
surrounding the bundle, due to the potential difference between the tip and bundle. This potential energy landscape is shown in Figure \ref{fig:regimes}e. This experiment differs from the TEM environment where plane waves are used; in the LEEPS setup, spherical waves emanating from a point source are deflected by the bi-prism. The deflection of the beams due to the bi-prism, calculated at 5 $\mu$Radian/V, is sufficient to cause the tune the overlap angle of the virtual sources and is convergent. As such, the overlap angle calculated here is represents a lower bound for the coherence angle of the original beam.

To describe the physics in our experimental setup, it's essential to note the interplay of  
three distinct types of waves (ignoring inelastically scattered and secondary waves) 
whose interplay gives rise to the pattern at the detector: 
\begin{enumerate}
\item the unperturbed reference wave from the nanotip; 
\item the elastically scattered wave from the sample (this gives rise to an 
in-line hologram, or Fresnel diffraction, when interfering with the reference wave,
and to a Young's double slit pattern, when two such waves scattered by the 
sample edges interfere); 
\item the wave deflected elastically by the electrostatic field surrounding 
the sample (such waves on each side of the sample giving rise to a bi-prism 
interference pattern). As Figure \ref{fig:regimes}e shows, equipotential contours near the 
grounded CNT bundle are highly curved, causing electron trajectories to be 
turned toward the axis. The grounded supporting membrane of the sample-holder is accounted for in this calculation, but not shown in this figure, being more distant.
\footnote{Electrostatic calculations were performed for the shown geometry using
the finite element method. Dirichlet boundary conditions were used for the tip, CNT and supporting membrane.
A custom mesh was employed to accurately capture the regions of high field.}
\end{enumerate}
 
Having laid out the principle effects at work, we can now review the presented data: at low magnification, such as Figure \ref{fig:branches} (a), when the tip is relatively far from the CNT bundle sample, most of the electron emission from the tip passes 
unperturbed to the detector.  A small fraction scatters off of the edges of the bundle,
causing Fresnel diffraction resulting in an in-line hologram.
As the magnification is progressively increased as in Figures
\ref{fig:branches} (c) through (f), two things happen which dramatically
change the pattern observed at the screen: 

(A) The field around the sample (scaling roughly as $U_{f}/ ln(d/R_{0})$) increases due 
to smaller tip-sample separation, $d$,  while the tip-sample bias $U_{f}$ changes little
($R_{0}$ is the local radius of curvature of the CNT bundle); the angle subtended by the bi-prism field encompasses the emission area of the tip and the deflection of the beam toward the axis increases.

(B) The reference wave (i) becomes increasingly blocked by the sample itself, 
while the unblocked portion is increasingly deflected by the bi-prism
effect, turning into wave of type (iii).

This latter effect becomes overwhelmingly dominant in cases of extreme magnification
such as in Figure  \ref{fig:branches} (f), where there is no visible trace left 
of the shadow image or holographic fringes, indicating that all radiation reaching
the detector has been deflected by the electrostatic field around the sample, and
only a small fraction of radiation will forward-scatter from the edges of
the sample. In this case, we have essentially a pure bi-prism experiment for a spherical 
electron wave emitted from a single atom. 
Recalling that fringe spacing is inversely related to source separation, it is observed that the relatively finely 
spaced fringes of \ref{fig:regimes}c are consistent with the more widely spaced virtual sources of a biprism 
and not consistent with the double-slit scattering by the two edges of the nanotube bundle.


These results taken together illustrate that carbon nanotubes
can be used to tailor the interference of the electron beam in LEEPS
\cite{georges2001imaging,prigent2000charge} and in this work can
be used to measure the coherence properties of the beam.  
Crucially, in the general case the fringe pattern is not purely 
sinusoidal because the pattern is not only due to the bi-prism; 
it also contains a hologram of the sample, which may or may not 
be distinguished in the total pattern, depending on the magnification
and the potential bias applied. Optimizing and reconstructing such holograms 
is a future goal.


The different fringe patterns discussed above can be seen distinctly by filtering
the data as shown in Figure \ref{fig:regimes} (a-c). The broader
set of fringes is the Fresnel diffraction pattern from scattering
off the bundle. The finer, sinusoidally-spaced fringes are due to the 
bi-prism effect described in equation \ref{eq:pattern} and illustrated in \ref{fig:regimes}c.
The two sets of fringes can be accentuated by filtering the images.
The finer fringes are accentuated by using a high-pass filter, as
in Figure \ref{fig:regimes}c. More details of the bi-prism 
effect are discussed in the supplementary information.

The coherence angle ($\alpha$ shown in Figure \ref{fig:regimes}) was
measured simply by calculating 

\[
\tan^{-1}(\frac{w}{2L}),
\]

where $w$ was the width of the pattern on the MCP and L was the tip-MCP
distance. The maximum coherence angle for these interference patterns
(as shown in Figure \ref{fig:coh_angle}) was 14.3 $\pm0.5^{\circ}$.
This represents a marked improvement over the previous state of the
art. 

A high coherence angle corresponds to high transverse components 
of electron wavevectors and this leads to an increase in
the resolution. In point-source holography this is captured in 
the virtual source size.
Using Eq. \ref{eq:vss} these measurements yield a virtual source
size \cite{scheinfein1993aberrations} of 1.7 $\pm0.6$ \AA\ \cite{spence1994electron}
for emission at 89.5 V and 1.2 nA. Alternatively, the upper-bound on resolution
of the LEEPS microscope can be estimated using the diffraction limit
in Eq. \ref{eq:diff_limit}. From this we expect a resolution of 2.6
$\pm$0.6 \AA. 

From the virtual source size, emission current and solid angle, all
measured in the same experiment, we calculate a brightness (normalized to
100 kV) of 7.9$\times10{}^{9}$ A/sr cm$^{2}$. In terms of reduced brightness, 
this is more than four times greater than
highest reported value in TEM \cite{kawasaki2000fine}. 
However, this value is less than that for carbon nanotube field emitters \cite{de2004brightness}, 
which is not surprising as the measurements reported here are 
based on a setup maximizing the coherence angle and not the degeneracy of the beam. 
We anticipate that by reducing the temperature, optimizing the extraction voltage, 
and further refinement of tip apex region,
the degeneracy (and the reduced brightness) 
of our beam can be increased to a value 
at least two orders of magnitude greater than conventional field emission tips \cite{silverman2008quantum}.

The reasons for the improvement in coherence angle reported here over that reported previous
is not entirely known. It may be simply the result of mechanical
and electrical stability of the LEEPS microscope. Great care was taken
to minimize vibrations of the tip relative to the sample. The instrument
has been operated in atom resolving scanning tunneling microscopy
mode, demonstrating that vibrations are $\backsim$0.1 \AA\ in magnitude.
This may allow the microscope to resolve higher coherence angles than
previously reported. Other factors may also contribute to the attainment
of higher coherence angles, such as the nature of the tip etching
process. The nanotip in this work is made by removing material from
the shank of the tip, rather than building a sharp protrusion upon
a broad base. This results in a tip with a higher aspect-ratio. The
electron emission is expected to be broader for a higher aspect-ratio
tip, since an emitter with a larger radius of curvature will focus
the emission into a narrow beam \cite{unnanotip,unnanotip3}. Lastly,
the nature of the surface of the tip itself may play a role. As the
tip is etched, a protective nitrogen-rich coating remains which modifies
the work function of the tip. This through a proximity effect, may
play a subtle role in tuning the emission character of the tip. Further
theoretical studies of these contributing factors are necessary to
fully understand the novel properties of the nitrogen etched nanotip.

Whereas no previously studied point source has displayed greater than
5.5 degrees coherent opening angle \cite{longchamp2012low}, corresponding
to approximately 1 nm \cite{stevens2009resolving} resolution for
a LEEPS microscope, the tip described here exhibits greater than 14
degrees opening angle, indicating the potential for $\sim$1.7 \AA\ resolution
and a transformative result for electron holography by the LEEPS method.
Such measurements are presently underway. 

The high overall 
brightness of 7.9$\times10{}^{9}$ A/sr cm$^{2}$ at room temperature
demonstrates the potential for broad applicability of this source.
The coherence and the brightness properties of the nitrogen etched nanotip
beam make it attractive for electron interferometry and holography both at
low and high energies and as well for production of degenerate fermionic
ensembles for applications requiring correlated quantum mechanical
behavior.  As the fabrication process is rather simple, the barrier to
wider application is low.

\begin{acknowledgments}
This research was funded by the Natural Sciences and Engineering Research
Council and Alberta Innovates - Technology Futures. The authors would
like to thank the machine shop staff in the Physics Department 
at the University of Alberta for their quality work. 

\bibliographystyle{plain}
\bibliography{bibs}

\pagebreak{}
\end{acknowledgments}

\part*{Supplementary information\label{part:Supplementary-information}}

\section{Comparison with other sources}

\begin{table}[h]
\begin{tabular}{|c|c|c|c|c|}
\hline 
\multicolumn{1}{|c|}{Reference} & Emission HWHM & Pattern HWHM & Geometry limited maximum angle & Coherence angle\tabularnewline

\hline 
\cite{golzhauser1998holographic} & - & 1.6$^{\circ}$ & $\sim$8$^{\circ}$ & $\sim$3$^{\circ}$\tabularnewline

\hline 
\cite{de2004brightness} & - & - & 1.6$^{\circ}$& 1.4$^{\circ}$ \tabularnewline
\hline
\cite{morin1996low,prigent2000charge,prigent2001charge} & - & - & 1.4$^{\circ}$ & -\tabularnewline
\hline 
\cite{oshima2005highly} & 3.3$^{\circ}$ & 3.3$^{\circ}$ & 4.3$^{\circ}$ & too large for detector\tabularnewline
\hline 
\cite{chang2009fully,tsong2013} & 2.3$^{\circ}$ & 2.3$^{\circ}$ & 12.8$^{\circ}$ & -\tabularnewline
\hline 
\cite{longchamp2012low} & - & - & NA=0.48, 30$^{\circ}$ & 5.5$^{\circ}$\tabularnewline
\hline 
This work & 7.25$^{\circ}$ & 7.25$^{\circ}$ & variable & 14.5$^{\circ}$\tabularnewline
\hline 

\end{tabular}

\caption{\label{tab:comp}A comparison of coherence angle from other works}
\end{table}

There are many previous measurements of coherence angle, a broad selection are shown in Table \ref{tab:comp}. Some of these quantities are quoted directly from the referenced papers, where in other cases they had to be calculated from information provided about the experimental apparatus. This is compiled in order to compare our work using the same figures of merit presented in the literature.

\section{Tip etching}

\begin{suppfigure}[h]
\includegraphics[width=2.8cm,height=2.8cm]{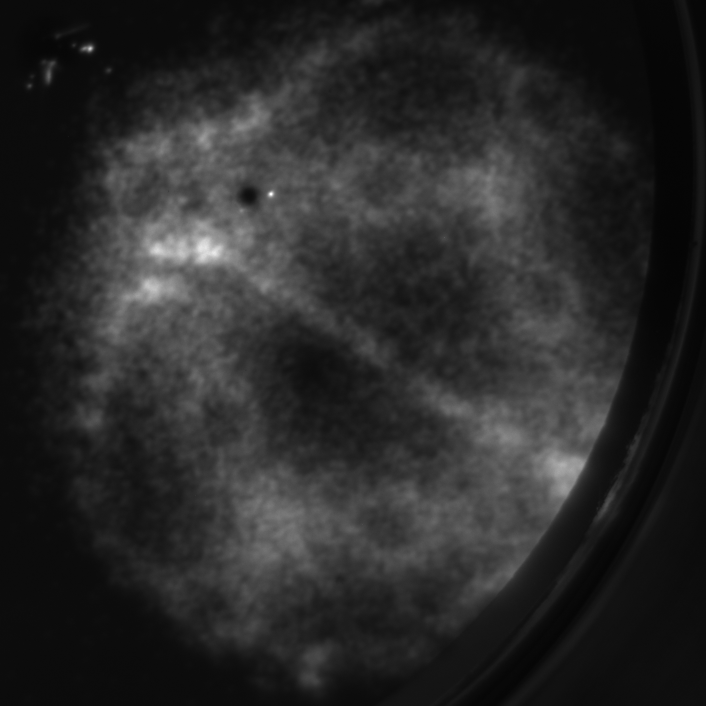}\includegraphics[width=2.8cm,height=2.8cm]{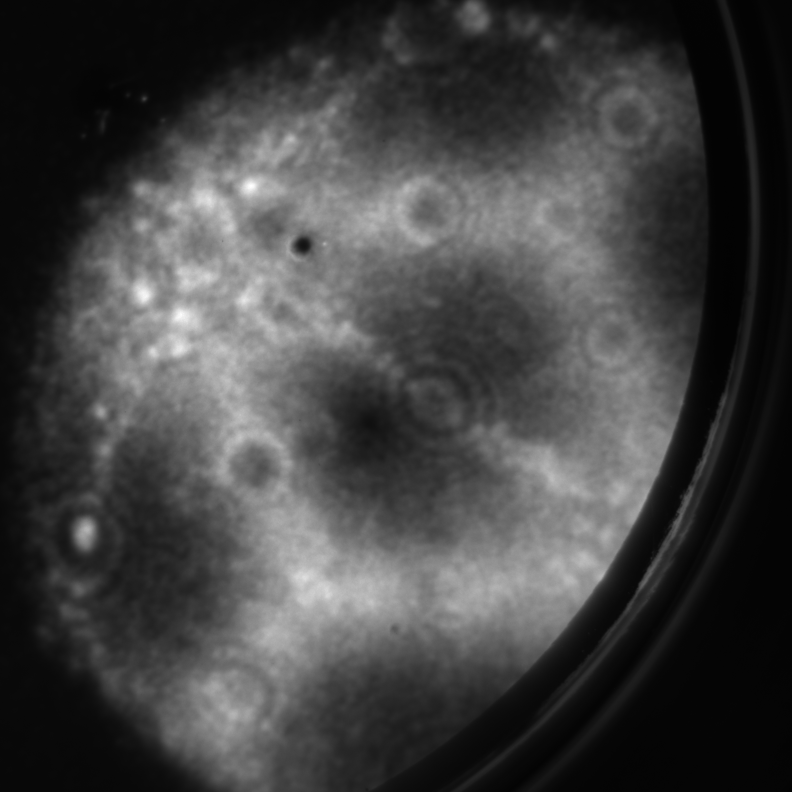}\includegraphics[width=2.8cm,height=2.8cm]{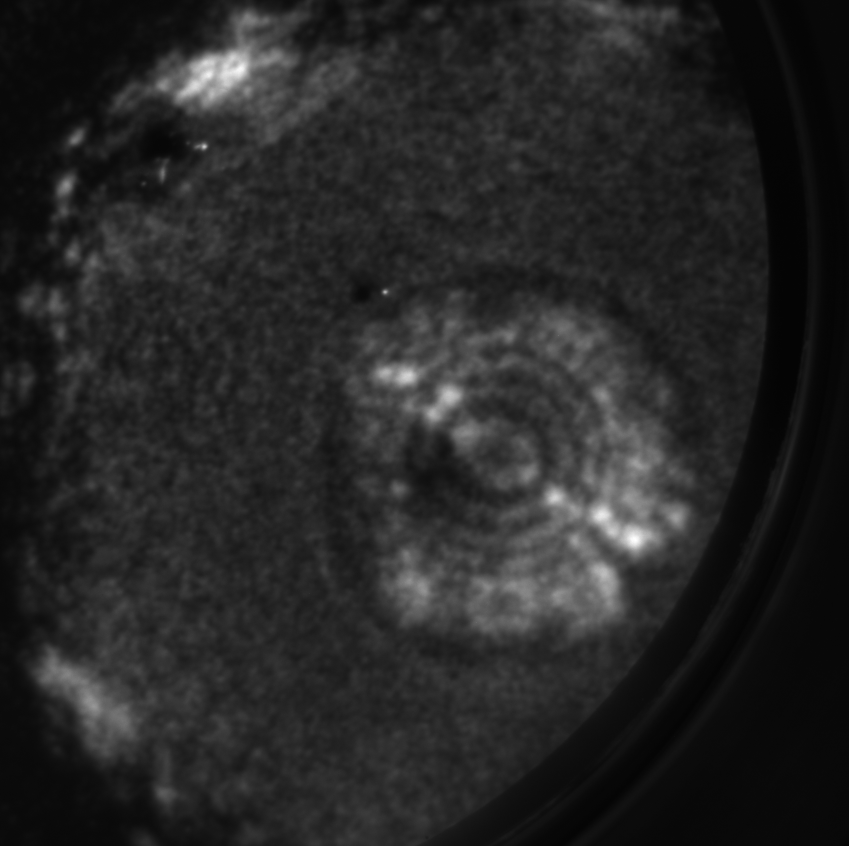}

\includegraphics[width=2.8cm,height=2.8cm]{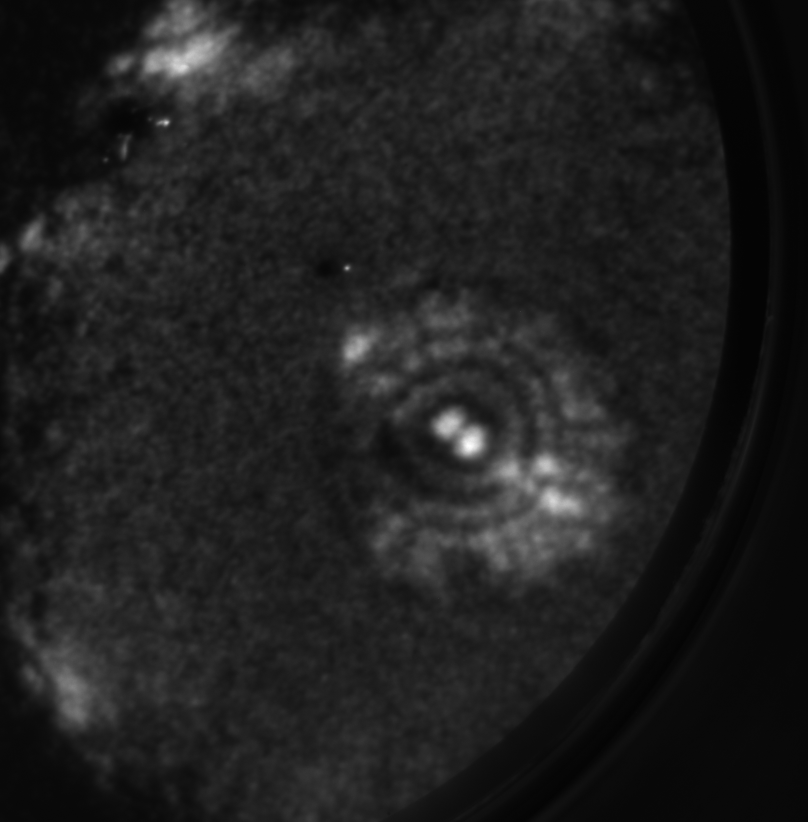}\includegraphics[width=2.8cm,height=2.8cm]{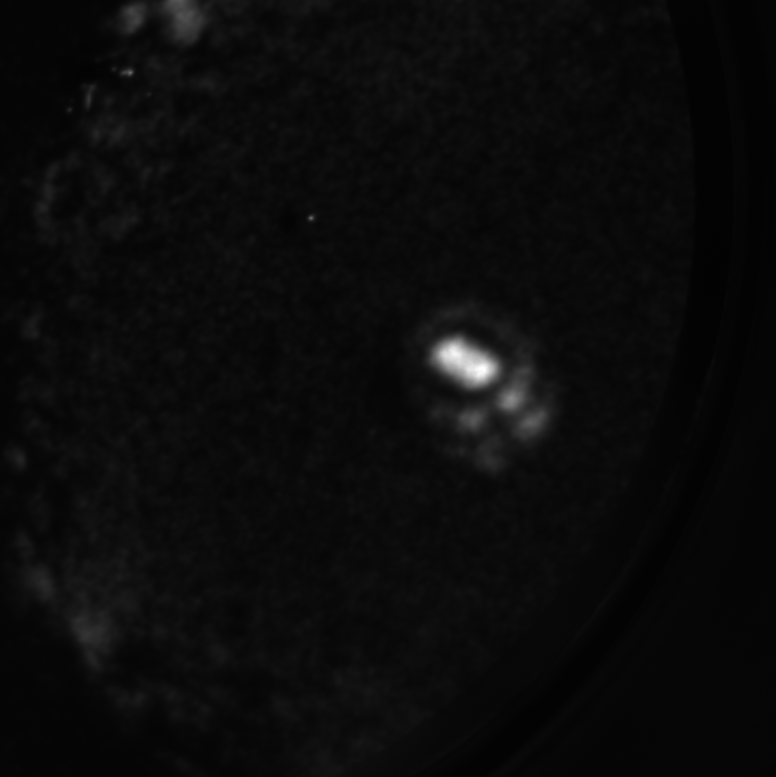}\includegraphics[width=2.8cm,height=2.8cm]{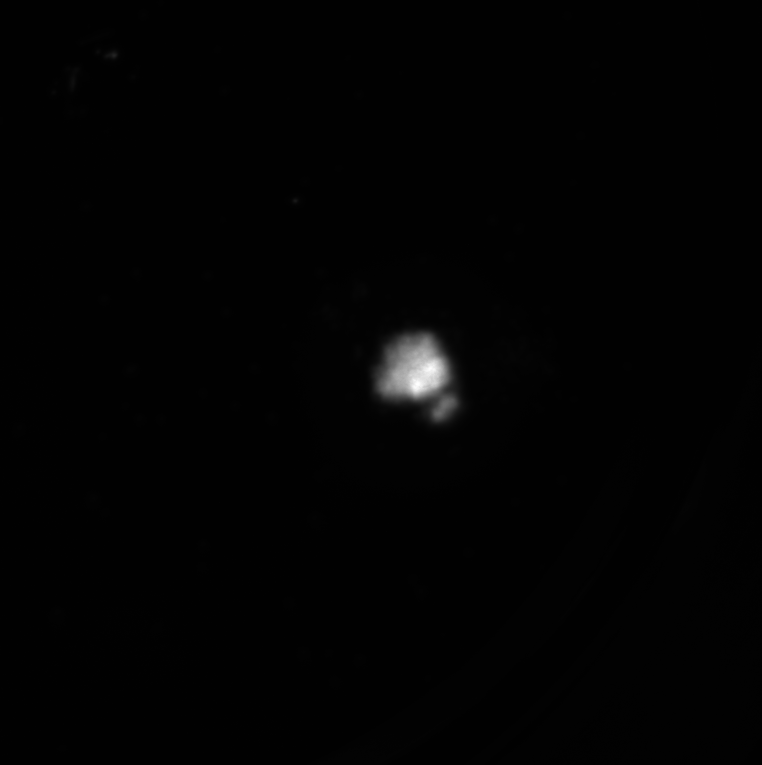}

\caption{FIM images showing the etching of the tip. Imaging voltages are 13.0
kV, 6.50 kV, 6.10 kV, 5.35kV, 5.05kV for each image, respectively}
\end{suppfigure}

To create such the tip used in the experiment, nitrogen gas is leaked
into the system under field-ion microscope (FIM) observation. Nitrogen
reacts with the tungsten, causing small protrusions. The field at
these protrusions is high enough to cause field-desorption - removing
the material. The high-fields used during the etching process inhibit
nitrogen from migrating near the apex. As a result the reaction and
field-desorption to occur preferentially at the shank of the tip.
What remains is a sharp apex with a higher aspect-ratio than single-atoms
tips fabricated using previous methods \cite{unnanotip,unnanotip3}\cite{pitters2012creation}. 

The tips in this work were etched using a specific recipe. Before
applying the field-assisted nitrogen technique, a relatively sharp
tip with a radius of curvature $>$10nm was first obtained by electrochemically
etching a polycrystalline tungsten wire in NaOH. The tip was then
loaded into a ultra-high vacuum (UHV) chamber with a base pressure
of 1x$10^{-10}$ Torr and placed in a LN$_{2}$ cooled FIM setup.
A high voltage was applied on the tip and a background pressure of
1x$10^{-5}$ Torr of helium was leaked into the chamber. A bias voltage
between 10-20 kV was necessary to obtain a clean, stable FIM image
of the tip. A background pressure of 2x$10^{-6}$ Torr of nitrogen
was then introduced into the chamber. The bias voltage was held constant
for 45 minutes. This allowed the material along the shank of the tip
outside of the FIM image, to be etched away. The voltage was then
decreased automatically at a rate of 1 V/s, until the bias reached
5.5 kV. When it was evident from the FIM image that etching had reached
a standstill, the voltage was lowered in 100 V increments at 1 V/s.
When the voltage reached 5 kV, the N$_{2}$ pressure was lowered to
8x$10^{-7}$ Torr and the voltage was lowered in 10 V increments until
the tip was sufficiently sharp, typically ending in a single atom.

\section{Interpretation of fringe patterns}

The electron trajectories are bent towards the optical axis by the
electric field resulting in an interference pattern analogous to that
of the double slit, where the slit separation is tuned by varying
the voltage between tip and bundle \cite{hasselbach2010progress,lichte1998gottfried}.
This is shown schematically in Figure \ref{fig:regimes}g. The
angular intensity of the interference pattern can be expressed as:
\begin{equation}
I(\alpha)=4I_{o}\cos^{2}\left(\frac{4\pi a\gamma_{o}U_{f}}{\lambda}\alpha\right),\label{eq:pattern}
\end{equation}
 where $a$ is the tip-bundle distance, $U_{f}$ is the relative bias
between tip and sample, $\lambda$ is the deBroglie wavelength of
the electron and $\gamma_{o}$ is the angular deflection per Volt
characteristic of the bi-prism. The bi-prism can be expressed as a
double slit with an effective slit separation of $D=a\gamma_{o}U_{f}$.
The pitch of the fringe pattern can be changed by tuning the voltage
of the tip relative to the bundle, as is shown in Supplementary Figure \ref{fig:biprism_graph}.
Practically, the tip-sample distance, extraction voltage and current
are related by the Fowler-Nordheim equation \cite{fowler1928electron}.
The tip can only be operated safely in a narrow window for beam current
ranging between 0.03 - 2 nA. This limits the amount we can explore
the dependence of the pattern on these variables. These finer fringe
patterns are only revealed at higher magnifications and are evident
in Figures \ref{fig:leeps_images}, \ref{fig:regimes}, and \ref{fig:coh_angle}.

\begin{suppfigure}
\includegraphics[height=.4\columnwidth]{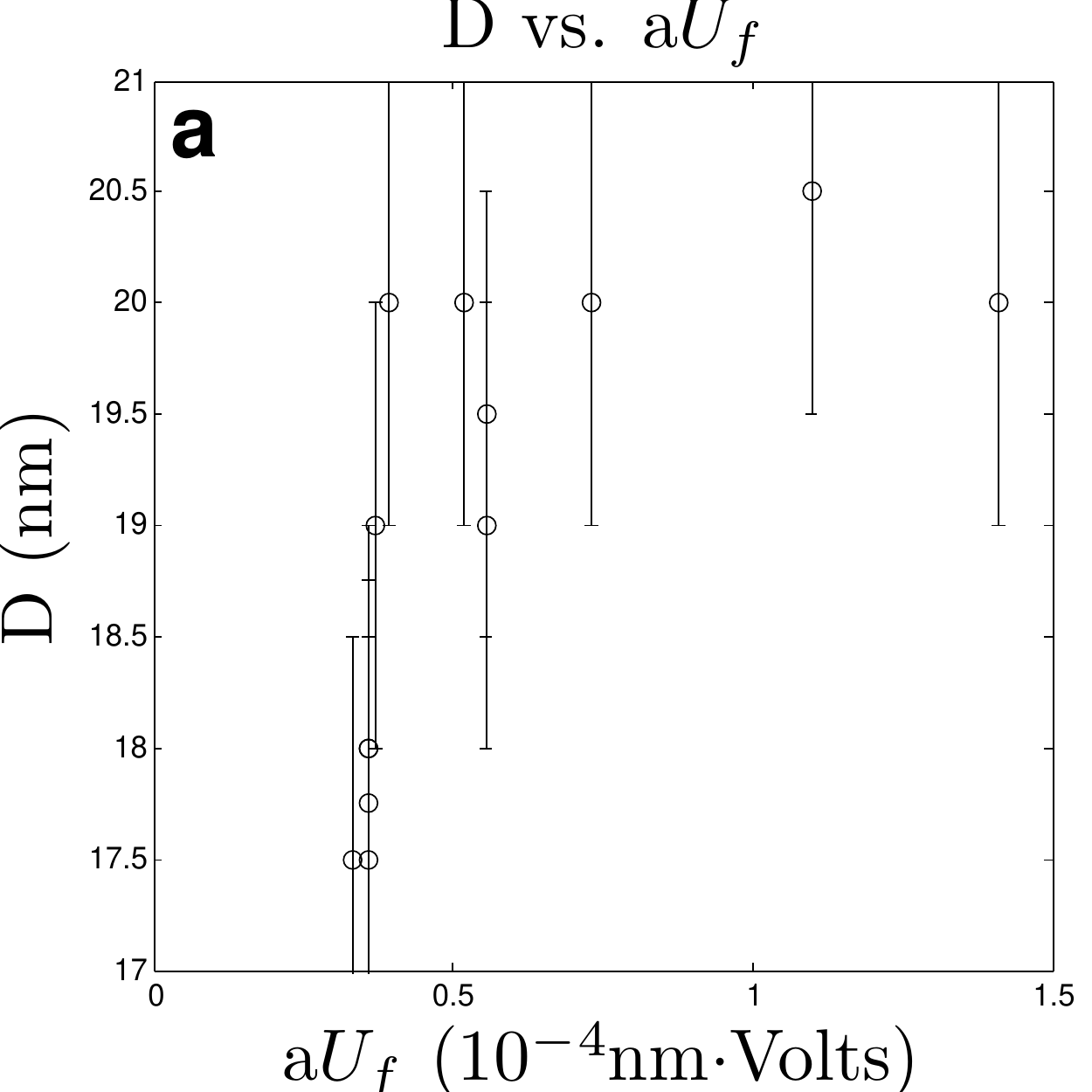}
\includegraphics[height=.4\columnwidth]{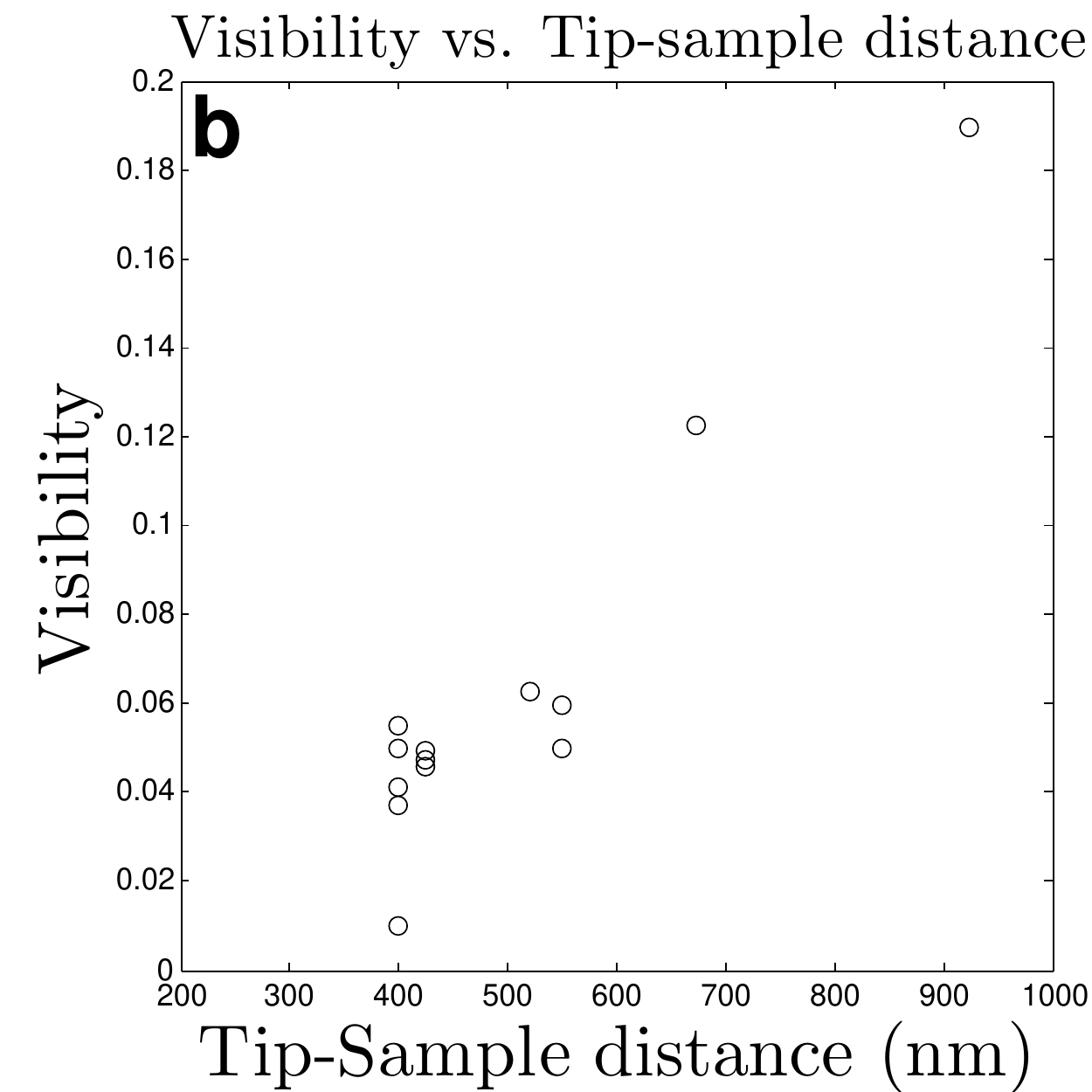}

\caption{The fine fringe pattern is generated by electrons that do not scatter off the nanotube bundle but instead are deflected by the
electric field surrounding it. This results in a bi-prism mode, analogous
to that of the double slit, where the effective slit separation can
be tuned by changing the bias between the tip and the bundle. This
is illustrated in \ref{fig:regimes}(g). As the field increases, so does the separation
of the virtual sources, resulting in a finer fringe pattern. (a):
demonstrates the dependence of the effective source separation on
tip-bundle bias $U_{f}$ and separation $a$. (b)The visibility of the fringe patterns falls off linearly
with tip-sample distance. This fits the conjecture that the reduced
visibility is simply to due obstruction of the coherence electron
wave by the nanotube bundle. }
\label{fig:biprism_graph}

\end{suppfigure}

It also must be noted that the visibility of the fringe patterns falls
off linearly as magnification increases as seen in  Supplementary Figure \ref{fig:biprism_graph}b.
While increased visibility has been previously seen with reduced temperature
of the emitter \cite{cho2004quantitative}, we doubt that the drop
off in visibility is due to a change in the coherence properties of
the source itself. This can be more simply explained by considering
the geometry of the experiment. As the magnification increases, the
tip is brought closer to the nanotube bundle, resulting in an increased
fraction of the beam being blocked by the tube. This results in
a greater fraction of electrons being absorbed or incoherently scattered
by the metallic nanotube bundle. This yields a smaller fraction of
coherent electron irradiation to contribute to the pattern, hence
the reduced visibility. This conjecture fits with the linear fall
off of visibility demonstrated in Supplementary Figure \ref{fig:biprism_graph}b: the solid angle (and therefore the number of electrons)
eclipsed by the bundle would vary linearly with tip-sample distance.
This lack of visibility due to the gross size of the sample also limits
the application of holographic reconstruction to this sample.





\end{document}